\documentclass{article} 
\usepackage{iclr2018_conference,times}
\usepackage{hyperref}
\usepackage{url}

\usepackage{amsmath}
\usepackage{graphicx}
\usepackage{caption}

\title{Emergence of grid-like representations by training recurrent neural networks to \\perform spatial localization}

\author{Christopher J. Cueva\thanks{equal contribution}, Xue-Xin Wei\footnotemark[1] \\
Columbia University\\
New York, NY 10027, USA \\
\texttt{\{ccueva,weixxpku\}@gmail.com} \\
}

\iclrfinalcopy 

\begin{document}

\maketitle

\begin{abstract}
Decades of research on the neural code underlying spatial navigation have revealed a diverse set of neural response properties. The Entorhinal Cortex (EC) of the mammalian brain contains a rich set of spatial correlates, including grid cells which encode space using tessellating patterns. However, the mechanisms and functional significance of these spatial representations remain largely mysterious. As a new way to understand these neural representations, we trained recurrent neural networks (RNNs) to perform navigation tasks in 2D arenas based on velocity inputs. Surprisingly, we find that grid-like spatial response patterns emerge in trained networks, along with units that exhibit other spatial correlates, including border cells and band-like cells. All these different functional types of neurons have been observed experimentally. The order of the emergence of grid-like and border cells is also consistent with observations from developmental studies. Together, our results suggest that grid cells, border cells and others as observed in EC may be a natural solution for representing space efficiently given the predominant recurrent connections in the neural circuits.
\end{abstract}

\section{Introduction}
Understanding the neural code in the brain has long been driven by studying feed-forward architectures, starting from Hubel and Wiesel's famous proposal on the origin of orientation selectivity in primary visual cortex~\citep{Hubel1962}. Inspired by the recent development in deep learning~\citep{Krizhevsky2012,Lecun2015,Hochreiter1997,Mnih2015}, there has been a burst of interest in applying deep feedforward models, in particular convolutional neural networks (CNN)~\citep{Lecun1998}, to study the sensory systems, which hierarchically extract useful features from sensory inputs~(see \emph{e.g.},~\citet{Yamins2014,Kriegeskorte2015,Kietzmann2017,Yamins2016}).

For more cognitive tasks, neural systems often need to maintain certain internal representations of relevant variables in the absence of external stimuli- a process that requires more than feature extraction. We will focus on spatial navigation, which typically requires the brain to maintain a representation of self-location and update it according to the animal's movements and landmarks of the environment. Physiological studies done in rodents and other mammals (including humans, non-human primates and bats) have revealed a variety of neural correlates of space in Hippocampus and Entorhinal Cortex (EC), including place cells~\citep{OKeefe1976}, grid cells~\citep{Fyhn2004,Hafting2005,Fyhn2008grid,Yartsev2011,Killian2012,Jacobs2013}, along with border cells~\citep{Solstad2008}, band-like cells~\citep{Krupic2012} and others (see Figure 1a). In particular, each grid cell only fires when the animal occupies a distinct set of physical locations, and strikingly these locations lie on a lattice.
The study of the neural underpinning of spatial cognition has provided an important window into how high-level cognitive functions are supported in the brain~\citep{Moser2008,Aranov2017}.

How might the spatial navigation task be solved using a network of neurons? Recurrent neural networks (RNNs)~\citep{Hochreiter1997,Graves2013,Oord2016,Theis2015,Gregor2015,Sussillo2015} seem particularly useful for these tasks. Indeed, recurrent-based continuous attractor networks have been one popular type of models proposed for the formation of grid cells~\citep{McNaughton2006,Burak2009,Couey2013} and place cells~\citep{Samsonovich1997}. 
Such models have provided valuable insights into one set of possible mechanisms that could support the formation of the grids. However, these models typically rely on fine-tuned connectivity patterns, in particular the models need a subtle yet systematic asymmetry in the connectivity pattern to move the attractor state according to the animal's own movement. The existence of such a specific 2D connectivity in rodent EC remains unclear.
Additionally, previous models have mainly focused on grid cells, while other types of responses that co-exist in the Entorhinal Cortex have been largely ignored. It would be useful to have a unified model that can simultaneously explain different types of neural responses in EC.

Motivated by these considerations, here we present an alternative modeling approach for understanding the representation of space in the neural system. Specifically, we trained a RNN to perform some spatial navigation tasks. By leveraging the recent development in RNN training and knowledge of the  navigation system in the brain, we show that training a RNN with biologically relevant constraints naturally gives rise to a variety of spatial response profiles as observed in EC, including grid-like responses. To our knowledge, this is the first study to show that grid-like responses could emerge from training a RNN to perform navigation. 

Our result implies that the neural representation in EC may be seen as a natural way for the brain to solve the navigation task efficiently~\citep{Wei2015}. More generally, it suggests that RNNs can be a powerful tool for understanding the neural mechanisms of certain high-level cognitive functions.

\begin{figure}[h]
 \centering
\includegraphics[keepaspectratio,width=0.8\linewidth]{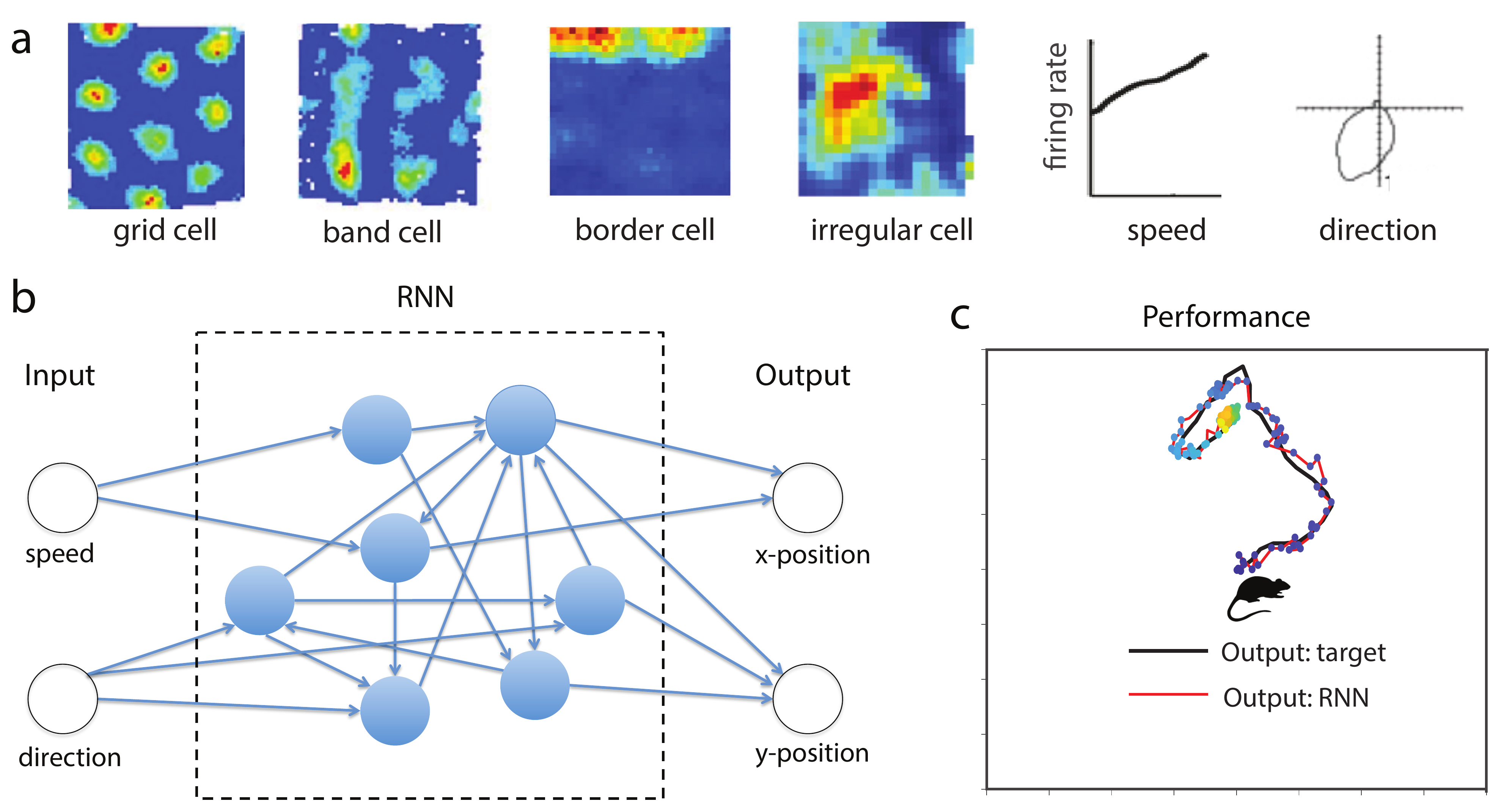}
 \caption{{\bf a)} Example neural data showing different kinds of neural correlates underlying spatial navigation in EC. All figures are replotted from previous publications. From left to right: a ``grid cell" recorded when an animal navigates in a square environment, replotted from~\cite{Krupic2012}, with the heat map representing the firing rate of this neuron as a function of the animal's location (red corresponds to high firing rate); a ``band-like" cell from \cite{Krupic2012}; a border cell from \cite{Solstad2008}; an irregular spatially tuned cell from \cite{Diehl2017}; a ``speed cell" from \cite{Kropff2015}, which exhibits roughly linear dependence on the rodent's running speed; a ``heading direction cell'' from \cite{Sargolini2006},  which shows systematic change of firing rate depending on animal's heading direction. {\bf b)} The network consists of $N=100$ recurrently connected units (or neurons) which receive two external inputs, representing the animal's speed and heading direction. The two outputs linearly weight the neurons in the RNN. The goal of training is to make the responses of the two output neurons accurately represent the animal's physical location.  {\bf c)} Typical trajectory after training. As shown, the output of the RNN can accurately, though not perfectly, track the animal's location during navigation.}
\end{figure}

\section{Model}


\subsection{Model description}
Our network model consists of a set of recurrently connected units ($N=100$). The dynamics of each unit in the network 
$u_i(t)$ is governed by the standard continuous-time RNN equation:
\begin{equation}
\tau \frac{dx_{i}(t)}{dt}  = -x_{i}(t) + \sum_{j=1}^{N}W^{\mathrm{rec}}_{ij}u_{j}(t)  + \sum_{k=1}^{N_{\mathrm{in}}}W^{\mathrm{in}}_{ik}I_{k}(t) + b_{i} + \xi_{i}(t) \label{eq:dynamics}
\end{equation}
for $i=1,\ldots,N$. The activity of each unit, $u_{i}(t)$, is related to the activation of that unit, $x_{i}(t)$, through a nonlinearity which in this study we take to be $u_{i}(t) = \tanh(x_{i}(t))$. Each unit receives input from other units through the recurrent weight matrix $W^{\mathrm{rec}}$ and also receives external input, $I(t)$, that enters the network through the weight matrix $W^{\mathrm{in}}$. Each unit has two sources of bias, $b_{i}$ which is learned and $\xi_{i}(t)$ which represents noise intrinsic to the network and is taken to be Gaussian with zero mean and constant variance.  The network was simulated using the Euler method for $T=500$ timesteps of duration $\tau/10$.

To perform a 2D navigation task with the RNN, we linearly combine the firing rates of units in the network to estimate the current location of the animal. The responses of the two linear readout neurons, $y_{1}(t)$ and $y_{2}(t)$, are given by the following equation: 
\begin{equation}
y_{j}(t) = \sum_{i=1}^{N}W^{\mathrm{out}}_{ji}u_{i}(t) \label{eq:output}
\end{equation}

\subsection{Input to the network}
The network inputs and outputs were inspired by simple spatial navigation tasks in 2D open environments. The task resembles dead-reckoning (sometimes referred to as path integration), which is ethologically relevant for many animal species~\citep{Darwin1873,Mittelstaedt1980,Etienne2004,McNaughton2006}.
To be more specific, the inputs to the network were the animal's speed and direction at each time step. Experimentally, it has been shown that the velocity signals exist in EC~\citep{Sargolini2006,Kropff2015,Hinman2016}, and there is also evidence that such signals are necessary for grid formation~\citep{WinterClark2015,Winter2015}.

Throughout the paper, we adopt the common assumption that the head direction of the animal coincides with the actual moving direction. The outputs were the x- and y-coordinates of the integrated position. The direction of the animal is modeled by modified Brownian motion to increase the probability of straight-runs, in order to be consistent with the typical rodent's behavior in an open environment. The usage of such simple movement statistics has the advantage of having full control of the simulated trajectories. However, for future work it would be very interesting to test the model using different animals' real movement trajectories to see how the results might change.

Special care is taken when the animal is close to the boundary. The boundary of the environment will affect the statistics of the movement, as the animal cannot cross the boundary. This fact was reflected in the model by re-sampling the angular input variable until the input angle did not lead the animal outside the boundary.  In the simulations shown below, the animal always starts from the center of the arena, but we verified that the results are insensitive to the starting locations.

\subsection{Training}
We optimized the network parameters $W^{\mathrm{rec}}$, $W^{\mathrm{in}}$, $b$ and $W^{\mathrm{out}}$ to minimize the squared error in equation (\ref{eq:error}) between target x- and y-coordinates from a two dimensional navigation task (performed in rectangular, hexagonal, and triangular arenas) and the network outputs generated according to equation (\ref{eq:output}).
\begin{equation}
E = \frac{1}{MTN_{\mathrm{out}}}\sum_{m,t,j=1}^{M,T,N_{\mathrm{out}}} (y_{j}(t,m) - y^{\mathrm{target}}_{j}(t,m))^{2} \label{eq:error}
\end{equation}
Parameters were updated with the Hessian-free algorithm~\citep{Martens2011} using mini-batches of size $M=500$ trials. In addition to minimizing the error function in equation (\ref{eq:error}) we regularized the input and output weights according to equation (\ref{eq:regL2}) and the squared firing rates of the units (referred to as metabolic cost) according to equation (\ref{eq:regfiring}). In sum, the training aims to minimize a loss function, that consists of the error of the animal, the metabolic cost, and a penalty for large network parameters.

\begin{align}
R_{L2}= &\frac{1}{NN_{\mathrm{in}}} \sum_{i,j=1}^{N,N_{\mathrm{in}}}(W^{\mathrm{in}}_{ij})^{2} + \frac{1}{NN_{\mathrm{out}}} \sum_{i,j=1}^{N_{\mathrm{out}},N}(W^{\mathrm{out}}_{ij})^{2} \label{eq:regL2} \\
R_{FR}= &\frac{1}{NTM}\sum_{i,t,m=1}^{N,T,M}u_{i}(t,m)^{2} \label{eq:regfiring}
\end{align}

We find that the results are qualitatively insensitive to the initialization schemes used for the recurrent weight matrix $W^{\mathrm{rec}}$. For the results presented in this paper, simulations in the hexagonal environment were obtained by initializing the elements of $W^{\mathrm{rec}}$ to be zero mean Gaussian random variables with variance $1.5^{2}/N$, and simulations in the square and triangular environments were initialized with an orthogonal $W^{\mathrm{rec}}$~\citep{Saxe2014}. We initialized the bias $b$ and output weights $W^{\mathrm{out}}$ to be zero. The elements of $W^{\mathrm{in}}$ were zero mean Gaussian variables with variance $1/N_{\mathrm{in}}$.

\begin{figure}[ht]
 \centering
\includegraphics[keepaspectratio,width=0.9\linewidth]{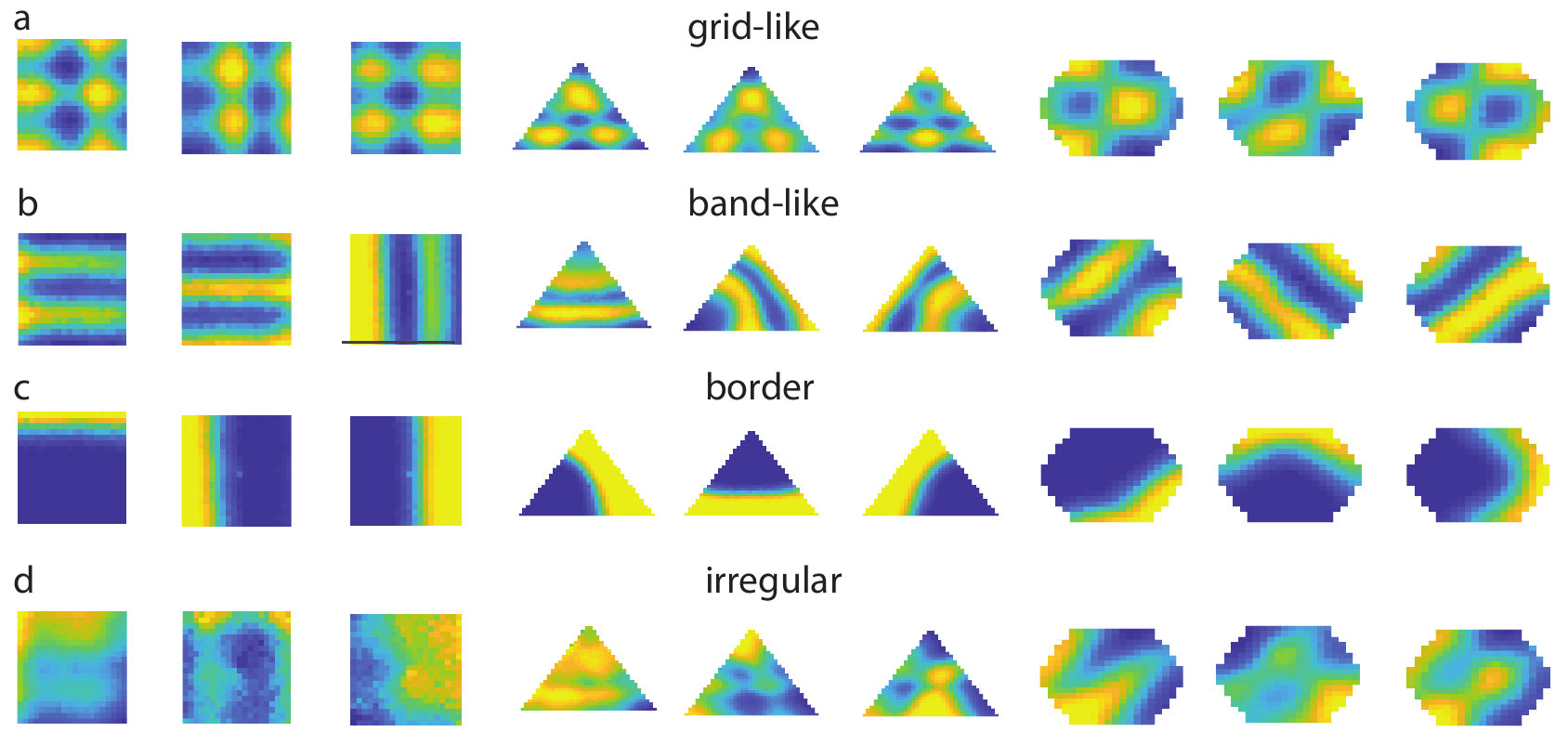}
 \caption{Different types of spatial selective responses of units in the trained RNN. Example simulation results for three different environments (square, triangular, hexagon) are presented. Blue (yellow) represents low (high) activity. {\bf a)} Grid-like responses. {\bf b)} Band-like responses; {\bf c)} Border-related responses; {\bf d)} Spatially irregular responses. These responses can be spatially selective but they do not form a regular pattern defined in the conventional sense.  }
\end{figure}

\section{Results}

We run simulation experiments in arenas with different boundary shapes, including square, triangular and hexagonal.
Figure 1c shows a typical example of the model performance after training; the network (red trace) accurately tracks the animal's actual path (black).

\subsection{Tuning properties of the model neurons}

We are mostly interested in what kind of representation the RNN has learned to solve this navigation task, and whether such a representation resembles the response properties of neurons in EC~\citep{Moser2008}.

\subsubsection{Spatial tuning}

To test whether the trained RNN developed location-selective representations, we plot individual neurons' mean activity level as a function of the animal's location during spatial exploration. Note that these average response profiles should not be confused with the linear filters typically shown in feedforward networks.
Surprisingly, we find neurons in the trained RNN show a range of interesting spatial response profiles. Examination of these response profiles suggests they can be classified into distinct functional types. Importantly, as we will show, these distinct spatial response profiles can be mapped naturally to known physiology in EC.  The spatial responses of all units in trained networks are shown in the Appendix.

{\bf Grid-like responses}
Most interestingly, we find some of the units in the RNN exhibit clear grid-like responses (Figure 2a). These firing patterns typically exhibit multiple firing fields, with each firing field exhibiting roughly circular symmetric or ellipse shape. Furthermore, the firing fields are highly structured, \emph{i.e.}, when combined, are arranged on a regular lattice. 
Furthermore, the structure of the response lattice depends on the shape of the boundary. In particular, training the network to perform self-localization in a square environment tends to give rectangular grids. In hexagonal and triangular environments, the grids are closer to triangular.

Experimentally, it is shown that (medial) EC contains so-called grid cells which exhibit multiple firing fields that lie on a regular grid~\citep{Fyhn2004, Hafting2005}.
The grid-like firing patterns in our simulation are reminiscent of the grid cells in rodents and other mammals. However, we also notice that the the grid-like model responses typically exhibit few periods, not as many as experimental data (see Figure 1a). It is possible that using a larger network might reveal finer grid-patterns in our model. Nonetheless, it is surprising that the gird-like spatial representations can develop in our model, given there is no periodicity in the input. Another potential concern is that, experimentally it is reported that the grids are often on the corners of a triangular lattice~\citep{Hafting2005} even in square environments (see Figure 1a), though the grids are somewhat influenced by the shape of the environment. However, the rats in these experiments presumable had spatial experience in other environments with various boundary shapes. Experimentally, it would be interesting to see if grid cells would lie on a square lattice instead if the rats are raised in a single square environment - a situation we are simulating here.

{\bf Border responses}
Many neurons in the RNN exhibit selectivity to the boundary (Figure 2c). Typically, they only encode a portion of the boundary, e.g. one piece of wall in a square shaped environment. Such properties are similar to the border cells discovered in rodent EC~\citep{Solstad2008, Savelli2008, Lever2009}. Experimentally, border cells mainly fire along one piece of wall, although some have been observed to fire along multiple borders or along the whole boundary of the environment; interestingly, these multi-border responses were also observed in some RNN models. Currently, it is unclear how the boundary-like response profiles emerge~\citep{Solstad2008,Savelli2008, Lever2009}. Our model points to the possibility that the border cells may emerge without the presence of tactile cues. Furthermore, it suggests that border cell formation may be related to the movement statistics of the animals, i.e. due to the asymmetry of the movement statistics along the boundary. 

{\bf Band-like responses}
Interestingly, some neurons in the RNN exhibit band-like responses (Figure 2b). In most of our simulations, these bands tend to be parallel to one of the boundaries. For some of the units, one of the bands overlaps the boundary, but for others, that is not the case. Experimentally, neurons with periodic-like firing patterns have been recently reported in rodent EC. In one study, it has been reported that a substantial portion of cells in EC exhibit band-like firing characteristics~\citep{Krupic2012}. However, we note that based on the reported data in \citet{Krupic2012}, the band pattern is not as clear as in our model.

{\bf Spatially-stable but non-regular responses}
Besides the units described above, most of the remaining units also exhibit stable spatial responses, but they do not belong to the above categories. These response profiles can exhibit either one large irregular firing field; or multiple circular firing fields, but these firing fields do not show a regular pattern.
Experimentally these types of cells have also been observed. In fact, it is recently reported that the non-grid spatial cells constitute a large portion of the neurons in Layer II and III of rodent EC~\citep{Diehl2017}. 
\begin{figure}[h]
 \centering
\includegraphics[keepaspectratio,width=0.9\linewidth]{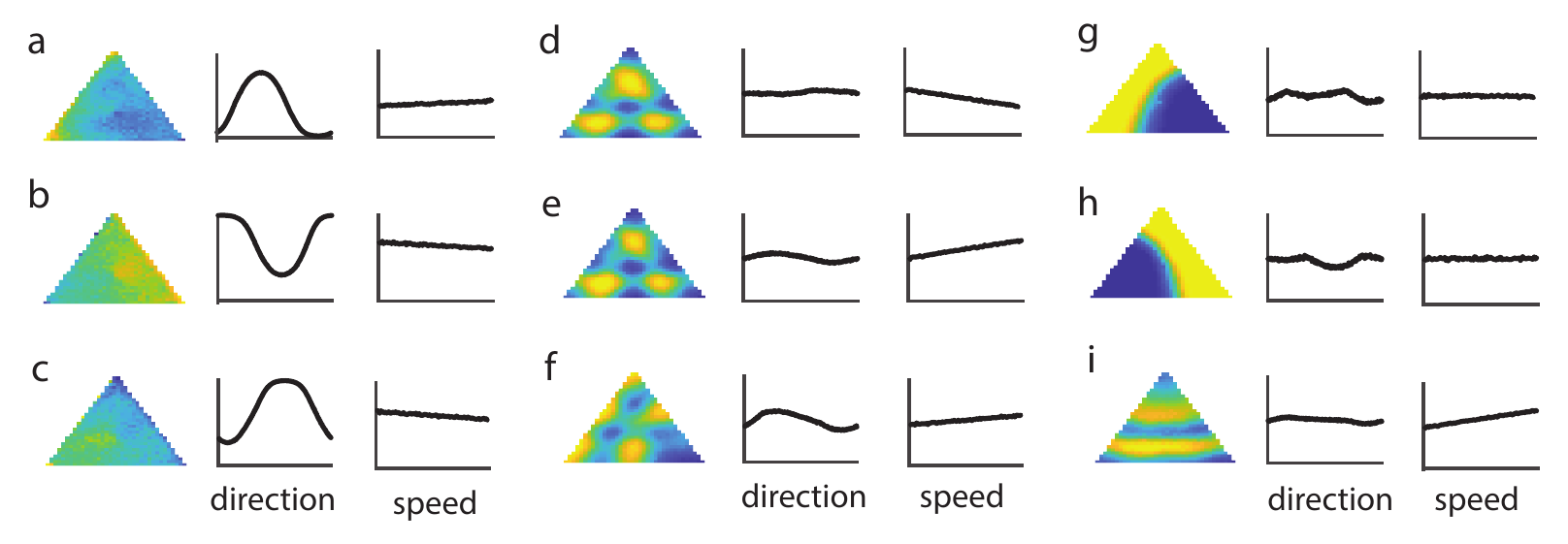}
 \caption{Direction tuning and speed tuning for nine example units in an RNN trained in a triangular arena. For each unit, we show the spatial tuning, (head) directional tuning, speed tuning respectively, from left to right. {\bf a,b,c)} The three model neurons show strong directional tuning, but the spatial tuning is weak and irregular. The three neurons also exhibit linear speed tuning. {\bf d,e,f)} The three neurons exhibit grid-like firing patterns, and clear speed tuning. The strength of their direction tuning differ. {\bf g,h)} Border cells exhibit weak and a bit complex directional tuning and almost no speed tuning. {\bf i)} This band cell shows weak directional tuning, but strong speed tuning.  }
\end{figure}

\subsubsection{Speed tuning and head direction tuning}

{\bf Speed tuning} We next ask how neurons in the RNN are tuned to the inputs. Many of the model neurons exhibit linear responses to the running speed of the animal, while some neurons show no selectivity to speed, as suggested by the near-flat response functions. Example response profiles are shown in Figure 3.  Interestingly, we observe that the model border cells tend to have almost zero speed-tuning (e.g., see Figure 3g,h).

{\bf Head direction tuning} A substantial portion of the model neurons show direction tuning. There are a diversity of direction tuning profiles, both in terms of the strength of the tuning and their preferred direction.  Example tuning curves are shown in Figure 3, and the direction tuning curves of a complete population are shown in the Appendix. Interestingly, in general model neurons which show the strongest head direction tuning do not show a clear spatial firing pattern (see Figure 3a,b,c). This suggests that there are a group of neurons which are mostly responsible for encoding the direction. We also notice that neurons with clear grid-like firing can exhibit a variety of direction tuning strengths, from weak to strong (Figure 3d,e,f). In the Appendix, we quantify the relation between these different tuning properties at the whole population level, which show somewhat complex dependence. 

Experimentally, the heading direction tuning in EC is well-known, \emph{e.g.}, \citet{Sargolini2006}. Both the grid and non-grid cells in EC exhibit head direction tuning~\citep{Sargolini2006}. 
Furthermore, the linear speed dependence of the model neurons is similar to the properties of speed cells reported recently in EC~\citep{Kropff2015}. Our result is also consistent with another recent study reporting that the majority of neurons in EC exhibit some amount of speed tuning~\citep{Hinman2016}. It remains an open question experimentally, at a population level, how different types of tuning characteristics in EC relate to each other.

\subsubsection{Development of the tuning properties}

We next investigate how the spatial response profiles evolve as learning/training progresses.
We report two main observations. First, neurons that fire selectively along the boundary typically emerge first. Second, the grid-like responses with finer spatial tuning patterns only emerge later in training. For visualization, we perform dimensionality reduction using the t-SNE algorithm~\citep{Maaten2008}. This algorithm embeds 100 model neurons during three phases of training (early, intermediate, and late) into a two-dimensional space according to the similarity of their temporal responses. Here the similarity metric is taken to be firing rate correlation. In this 2D space as shown in Figure 4a, border cell representations appear early and stably persist through the end of training. Furthermore, early during training all responses are similar to the border related responses. In contrast, grid-like cells typically undergo a substantial change in firing pattern during training before settling into their final grid-like representation (Figure 4b).

The developmental time line of the grid-like cells and border cells is roughly consistent with developmental studies in rodents.
Experimentally, it is known that border cells emerge earlier in development, and they exist at about 2 weeks after the rat is born~\citep{Bjerknes2014}. The grid cells mature only at about 4 weeks after birth~\citep{Langston2010,Wills2010,Bjerknes2014}. Furthermore, our simulations suggest the reason why border cells emerge earlier in development may be that computationally it is easier to wire-up a network that gives rise to border cell responses.

\begin{figure}[ht]
 \centering
\includegraphics[keepaspectratio,width=0.99\linewidth]{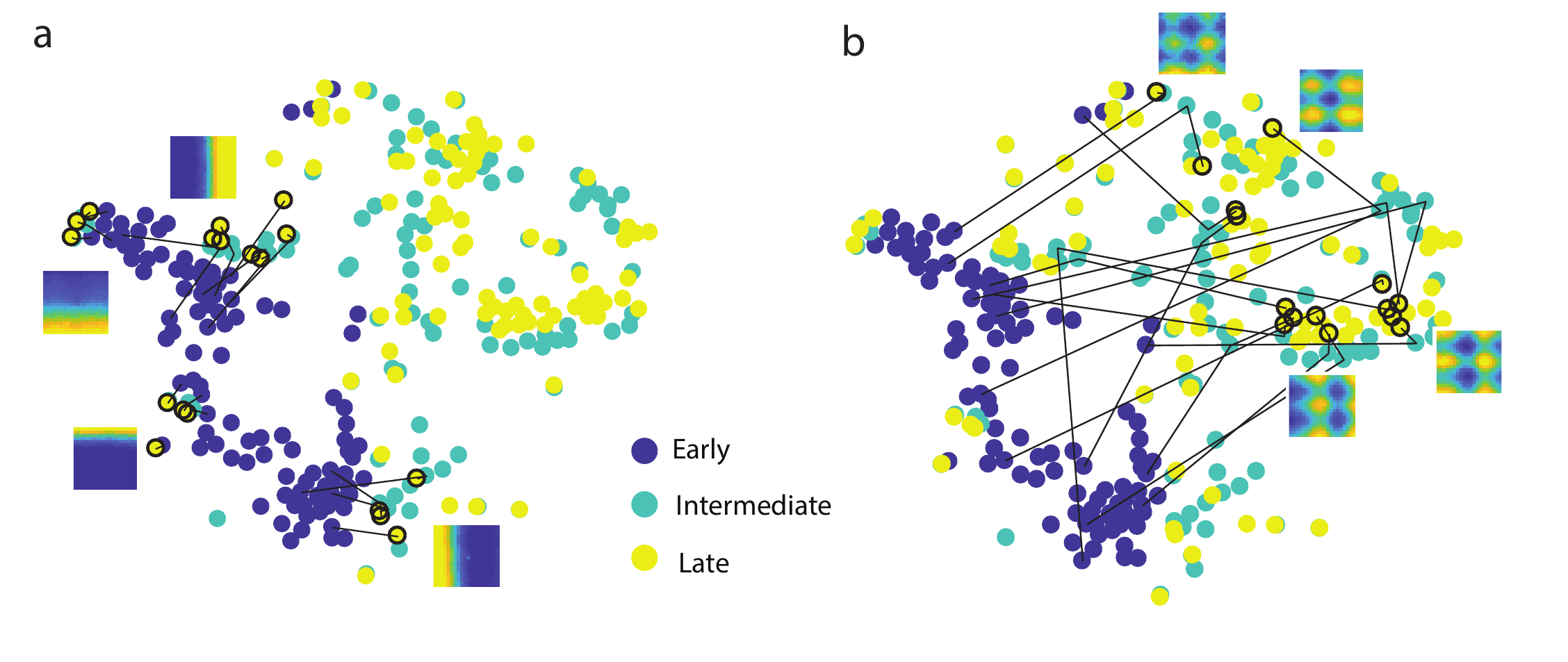}
 \caption{Development of border cells and grid-like cells. Early during training all responses are similar to the border related responses, and only as training continues do the grid-like cells emerge. We perform dimensionality reduction using the t-SNE algorithm on the firing rates of the neurons. Each dot represents one neuron ($N=100$), and the color represents different training stages (early/intermediate/late shown in blue/cyan/yellow). Each line shows the trajectory of a single highlighted neuron as its firing responses evolve during training. In panel {\bf a)}, we highlight the border representation. It appears there are four clusters of border cells, each responding to one wall of a square environment (spatial responses from four of these border cells are inset). These cells' response profiles appear early and stably persist through training, illustrated by the short distance they travel in this space. In {\bf b)}, we show that the neurons which eventually become grid cells initially have tuning profiles similar to the border cells but then change their tuning substantially during learning. As a natural consequence, they need to travel a long distance in this space between the early and late phase of the training. Spatial responses are shown for four of these grid-like cells during the late phase of training.}  
 \end{figure}

\subsection{The importance of regularization}

\begin{figure}[ht]
 \centering
\includegraphics[keepaspectratio,width=0.99\linewidth]{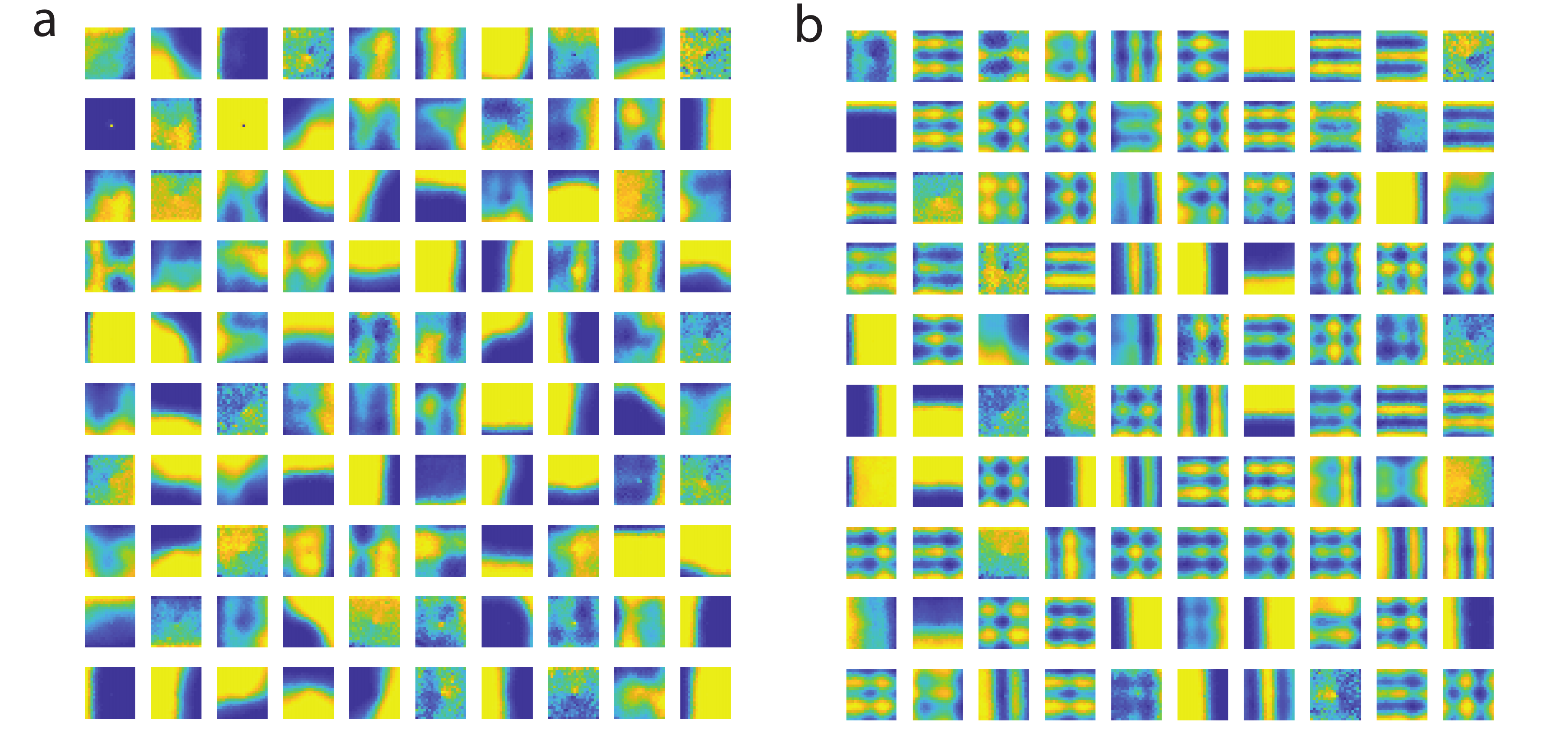}
 \caption{Complete set of spatial response profiles for 100 neurons in a RNN trained in a square environment. {\bf a)} Without proper regularization, complex and periodic spatial response patterns do not emerge. {\bf b)} With proper regularization, a rich set of periodic response patterns emerge, including grid-like responses. Regularization can also be adjusted to achieve spatial profiles intermediate between these two examples.}
\end{figure}

We find appropriate regularizations of the RNN to be crucial for the emergence of grid-like representations. We only observed grid-like representations when the network was encouraged to store information while perturbed by noise. This was accomplished by setting the speed input to zero, e.g. zero speed 90\% of the time, and adding Gaussian noise to the network ($\xi_{i}(t)$ in equation (\ref{eq:dynamics})); the precise method for setting the speed input to zero and the value of the noise variance is not crucial for our simulations to develop grid-like representations. The cost function which aims to capture the penalization on the metabolic cost of the neural activity also acts as an important regularization. Our simulations show that the grid-like representation did not emerge without this metabolic cost. In Figure 5, we show typical simulation results for a square environment, with and without proper metabolic regularization. In the Appendix, we illustrate the effect of regularization further, in particular the role of injecting noise into the RNN units.

Our results are consistent with the general notion on the importance of incorporating proper constraint for learning useful representations in neural networks~\citep{Bengio2013}. Furthermore, it suggests that, to learn a model with response properties similar to neural systems it may be necessary to incorporate the relevant constraints, \emph{e.g.}, noise and metabolic cost.

\subsection{Error correction around the boundary}

\begin{figure}[h]
 \centering
\includegraphics[keepaspectratio,width=0.9\linewidth]{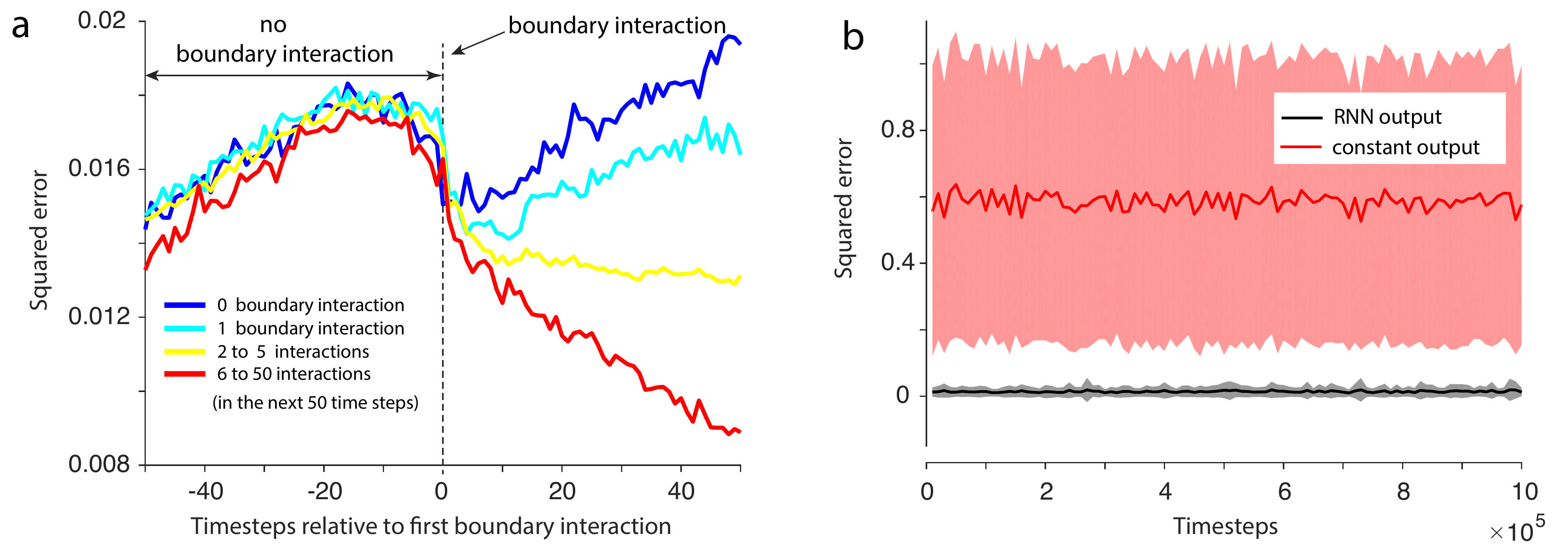}
 \caption{Error-correction happens at the boundary and the error is stable over time. At the boundary, the direction is resampled to avoid input velocities that lead to a path extending beyond the boundary of the environment. These changing input statistics at the boundary, termed a boundary interaction, are the only cue the RNN receives about the boundary.  We find that the RNN uses the boundary interactions to correct the accumulated error between the true integrated input and its prediction based on the linear readout of equation (2).  Panel {\bf a)}, the mean squared error increases when there are no boundary interactions, but then decreases after a boundary interaction, with more boundary interactions leading to greater error reduction. In the absence of further boundary interaction, the squared error would gradually increase again (blue curve) at roughly a constant rate. {\bf b)} The network was trained using mini-batches of 500 timesteps but has stable error over a duration at least four orders of magnitude larger. The error of the RNN output (mean and standard deviation shown in black, computed based on 10000 timesteps) is compared to the error that would be achieved by an RNN outputting the best constant values (red). }
\end{figure}

One natural question is whether the trained RNNs are able to perform localization when the path length exceeds the typical length used during training (500 steps), in particular given that noise in the network would gradually accumulate, leading to a decrease in localization performance. We test this by simulating paths that are several orders of magnitude longer. Somewhat surprisingly, we find the RNNs still perform well~(Figure 6b). In fact, the squared error (averaged over every 10000 steps) is stable. The spatial response profiles of individual units also remain stable. This implies that the RNNs have acquired intrinsic error-correction mechanisms during training.

As shown earlier, during training some of the RNN units develop boundary-related firing (Figure 2c), presumably by exploiting the change of input statistics around the boundary. We hypothesize that boundary interactions may enable error-correction through signals based on these boundary-related activities. Indeed, we find that boundary interactions can dramatically reduce the accumulated error~(Figure 6a). Figure 6a shows that, without boundary interactions, on average the squared error grows roughly linearly as expected, however, interactions with the boundaries substantially reduce the error, and more frequent boundary interactions can reduce the error further. Error-correction on grid cells via boundary interactions has been proposed~\citep{Hardcastle2015,Pollock2016}, however, we emphasize that the model proposed here develops the grid-like responses, boundary responses and the error-correction mechanisms all within the same neural network, thus potentially providing a unifying account of a diverse set of phenomena.

\section{Discussion}

In this paper, we trained RNNs to perform path integration (dead-reckoning) in 2D arenas. We found that after training RNNs with appropriate regularization, the model neurons exhibit a variety of spatial and velocity tuning profiles that match neurophysiology in EC. 
What's more, there is also similarity in terms of when these distinct neuron types emerge during training/development. The EC has long been thought to be involved in path integration and localization of the animal's location~\citep{Moser2008}. The general agreement between the different response properties in our model and the neurophysiology provide strong evidence supporting the hypothesis that the neural population in EC may provide an efficient code for representation self-locations based on the velocity input.

Recently, there has been increased interest in using complex neural network models to understand the neural code. But the focus has been on using feedforward architectures, in particular CNNs~\citep{Lecun1998}. Given the abundant recurrent connections in the brain, it seems a particularly fruitful avenue to take advantage of the recent development in RNNs to help with neuroscience questions~\citep{Mante2013,Song2016,Miconi2017,Sussillo2015}. Here, we only show one instance following this approach. However, the insight from this work could be general, and potentially useful for other cognitive functions as well.

The finding that metabolic constraints lead to the emergence of grid-like responses may be seen as conceptually related to the efficient coding hypothesis in visual processing \citep{Barlow1961}, in particular the seminal work on the emergence of the V1-like Gabor filters in a sparse coding model by \citet{Olshausen1996}. Indeed, our work is partly inspired by these results. While there are conceptual similarities, however, we should also note there are differences between the sparse coding work and ours. First, the sparsity constraint in sparse coding can be naturally viewed as a particular prior while in the context of the recurrent network, it is difficult to interpret that way. Second, the grid-like responses are not the most sparse solution one could imagine. In fact, they are still quite dense compared to a more spatially localized representation. Third, the grid-like patterns that emerged in our network are not filters based on the raw input, rather the velocity inputs need to be integrated first in order to encode spatial locations. Our work is also inspired by recent work using the efficient coding idea to explain the functional architecture of the grid cells~\citep{Wei2015}. It has been shown that efficient coding considerations could explain the particular set of grid scales observed in rodents \citep{Stensola2012}. However, in that work, the firing patterns of the neurons are assumed to have a lattice structure to start with. Furthermore, our work is related to the study by Sussillo and others \citep{Sussillo2015}, in which they show that regularization of RNN models are important for generating solutions that are similar to the neural activity observed in motor cortex. In Sussillo et al., a \emph{smoothness} constraint together with others lead to simple oscillatory neural dynamics that well matches the neural data. We have not incorporated a smoothness constraint into our network.

Additionally, we note that there are a few recent studies which use place cells as the input to generate grid cells~\citep{Dordek2016,Stachenfeld2016}, which are fundamentally different from our work. In these feedforward network models, the grid cells essentially perform dimensionality reduction based on the spatial input from place cells. However, the main issue with these models is that, it is unclear how place cells acquire spatial tuning in the first place. To the contrary, our model takes the animal's velocity as the input, and addresses the question of how the spatial tuning can be generated from such input, which are known to exist in EC~\citep{Sargolini2006,Kropff2015}. In another related study~\citep{Kanitscheider2016}, the authors train a RNN with LSTM units~\citep{Hochreiter1997} to perform different navigation tasks. However, no grid-like spatial firing patterns are reported.

Although our model shows a qualitative match to the neural responses observed in the EC, nonetheless it has several major limitations, with each offering interesting future research directions. First, the learning rule we use seems to be biologically implausible. We are interested in exploring how a more biologically plausible learning rule could give rise to similar results~\citep{Lillicrap2016,Miconi2017,Guerguiev2017}.
Second, the simulation results do not show a variety of spatial scales in grid-like cells. Experimentally, it is known that grid cells have multiple spatial scales, that scale geometrically with a ratio $1.4$~\citep{Stensola2012}, and this particular scale ratio is predicted by efficient coding of space \citep{Wei2015}. We are investigating how to modify the model to get a hierarchy of spatial scales, perhaps by incorporating more neurons or modifying the regularization.
Last but not least, we have focused on the representation produced by the trained RNN. An equally important set of questions concern how the networks actually support the generation of such a representation. As a preliminary effort, we have examined the connectivity patterns of the trained network, and they do not seem to resemble the connectivity patterns required by standard attractor network models. Maybe this should not be seen as too surprising. After all, the trained networks can produce a diverse set of neural responses, while the previous models only led to grid responses. It would be interesting for future work to systematically examine the questions related to the underlying mechanisms.

\section{Acknowledgements}
We thank members of the Center for Theoretical Neuroscience at Columbia University for useful discussions and three anonymous reviewers for constructive feedback. Research supported by NSF NeuroNex Award DBI-1707398 and NIH training grant 5T32NS064929 (CJC).

\bibliography{iclr2018_conference}
\bibliographystyle{iclr2018_conference}

\appendix

\newpage
\section{Triangular environment}
\begin{figure}[h]
 \centering
\includegraphics[keepaspectratio,width=0.48\linewidth]{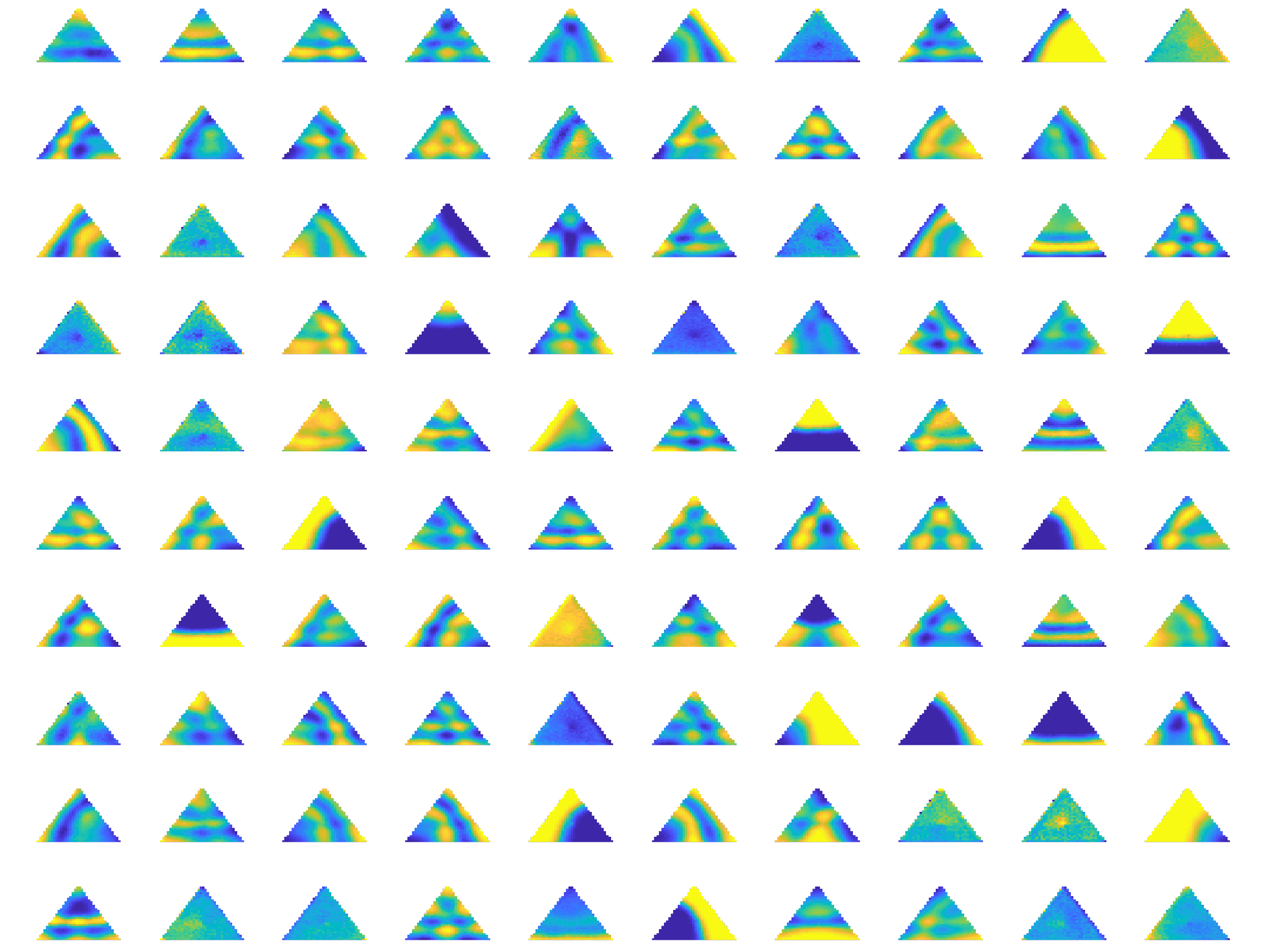}
\end{figure}

\begin{figure}[h]
\centering
\includegraphics[trim={0cm 9.5cm 0cm -2cm},clip,keepaspectratio,width=1.0\linewidth]{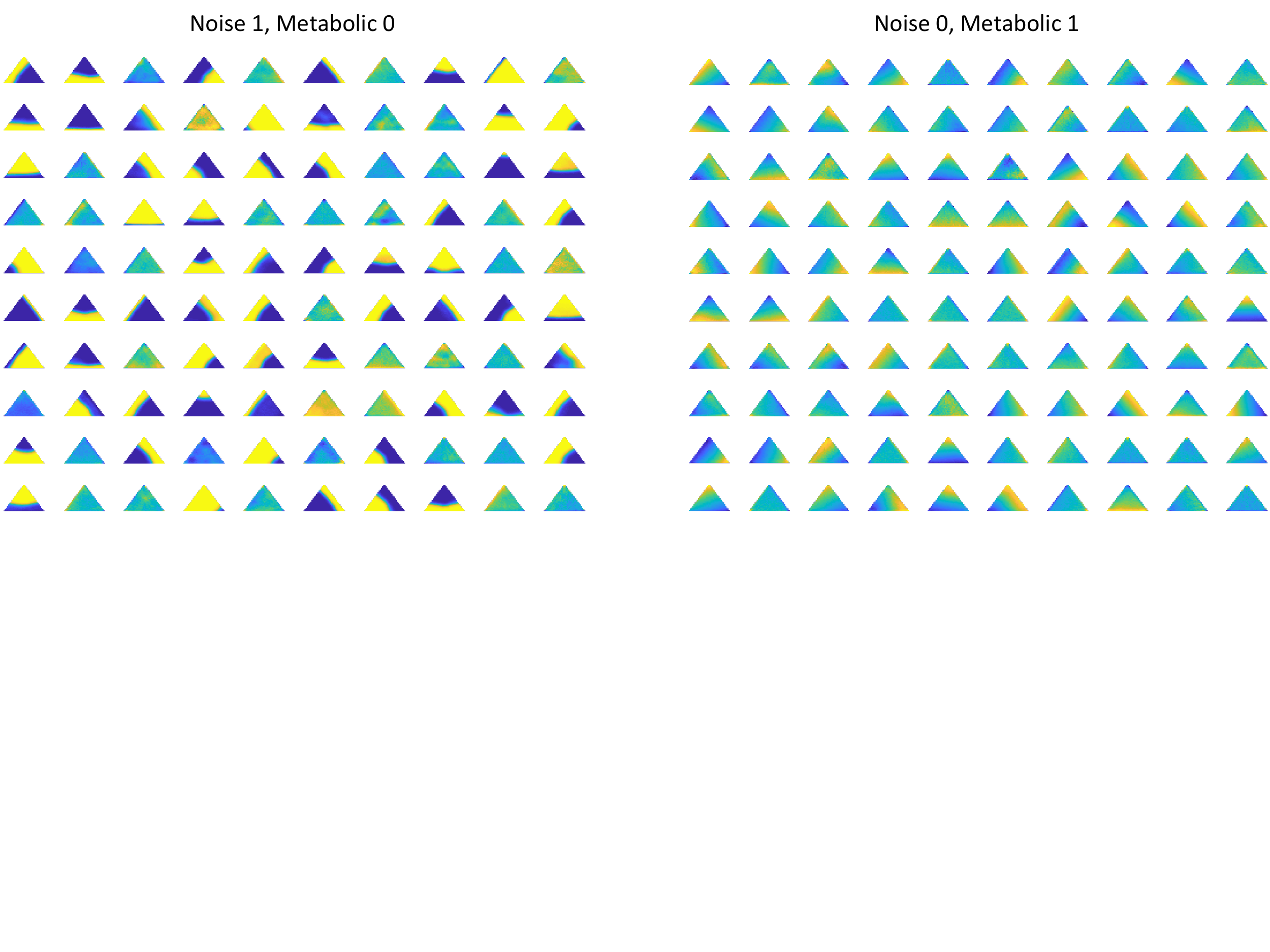}
\captionsetup{labelformat=empty}
\caption{Noise and metabolic cost are important for grid-like representations. The figure on the left shows the spatial responses for a network trained with noise and no metabolic cost. The figure on the right shows the spatial responses for a network trained with no noise and the metabolic cost.}
\end{figure}

\clearpage

\begin{figure}[h]
 \centering
\includegraphics[trim={0cm 0.75cm 0cm 0.9cm},clip,keepaspectratio,width=0.68\linewidth]{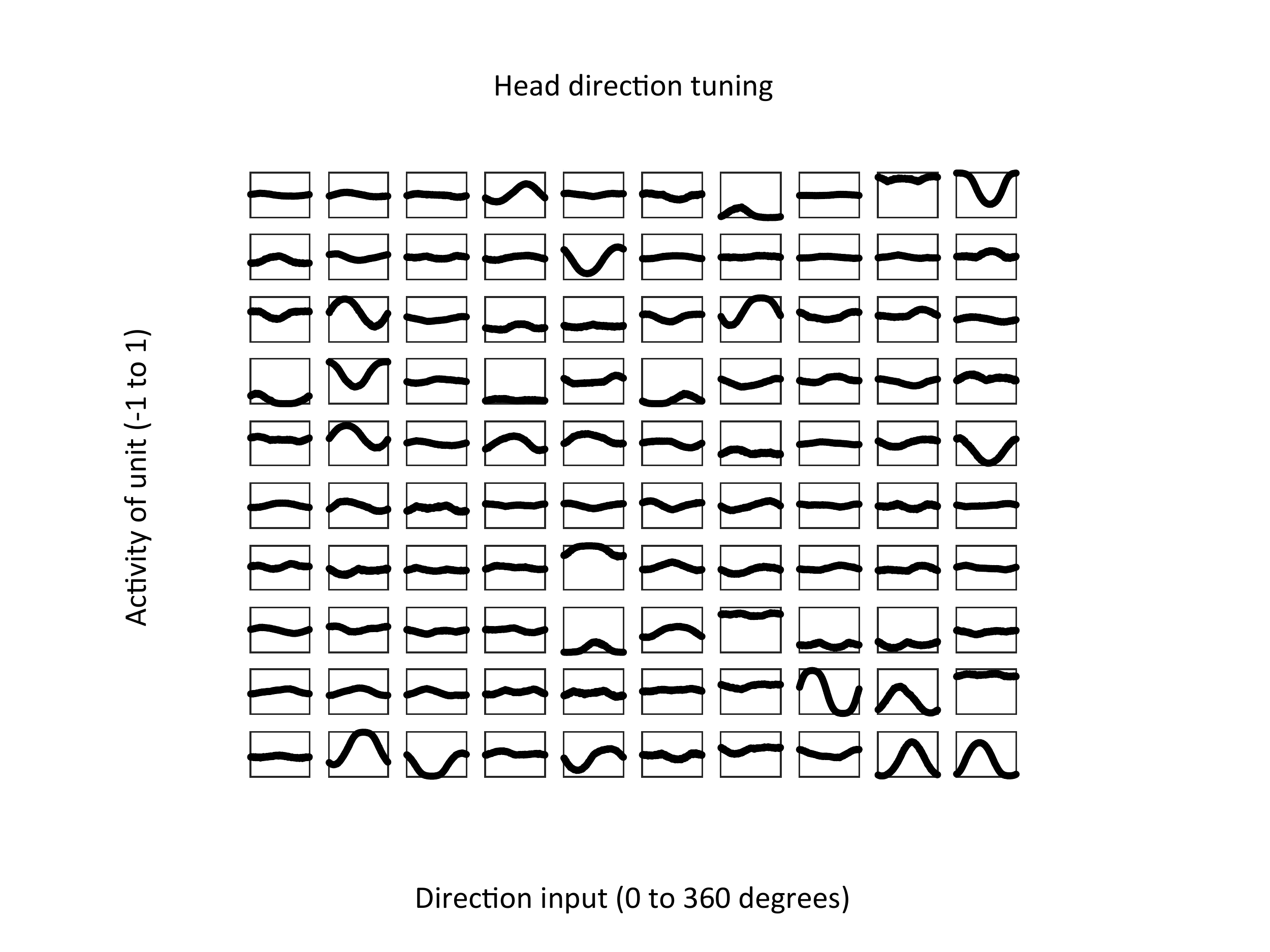}
\end{figure}
\begin{figure}[h]
 \centering
\includegraphics[trim={0cm 0.75cm 0cm 0.9cm},clip,keepaspectratio,width=0.68\linewidth]{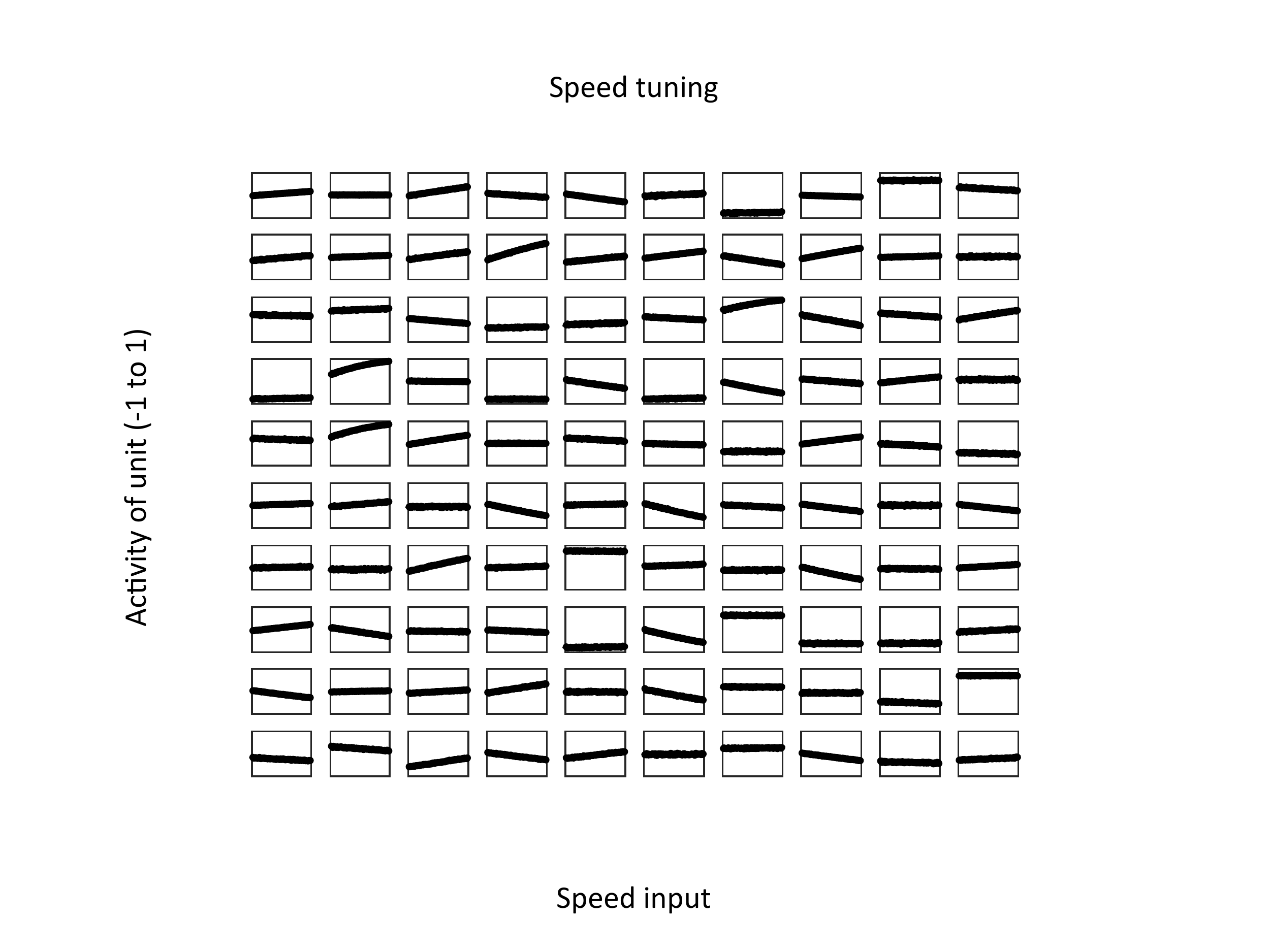}
\end{figure}

\clearpage
\section{Rectangular environment}
\begin{figure}[h]
 \centering
\includegraphics[keepaspectratio,width=0.48\linewidth]{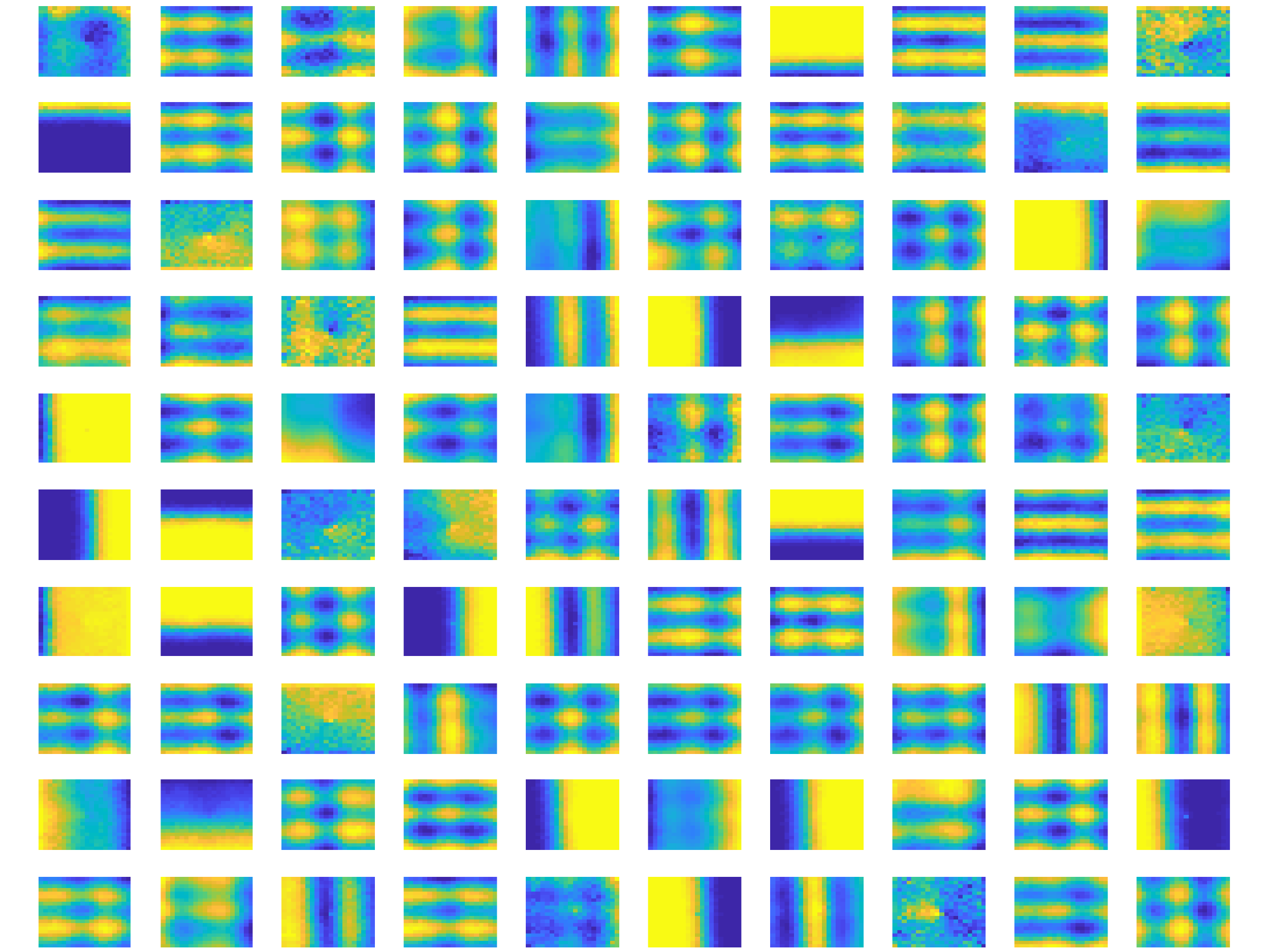}
\end{figure}
\begin{figure}[h]
 \centering
\includegraphics[keepaspectratio,width=0.68\linewidth]{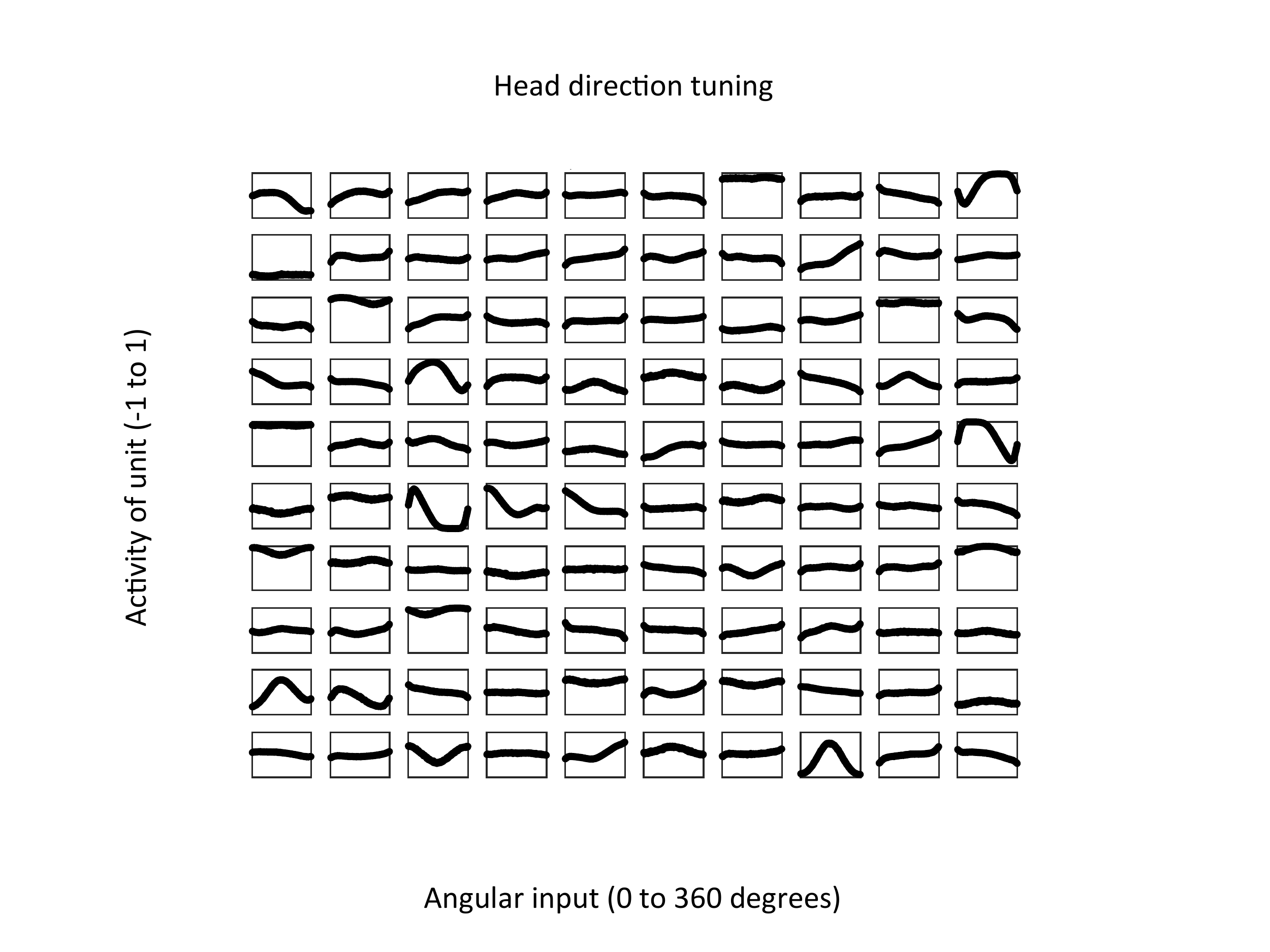}
\end{figure}
\begin{figure}[h]
 \centering
\includegraphics[keepaspectratio,width=0.68\linewidth]{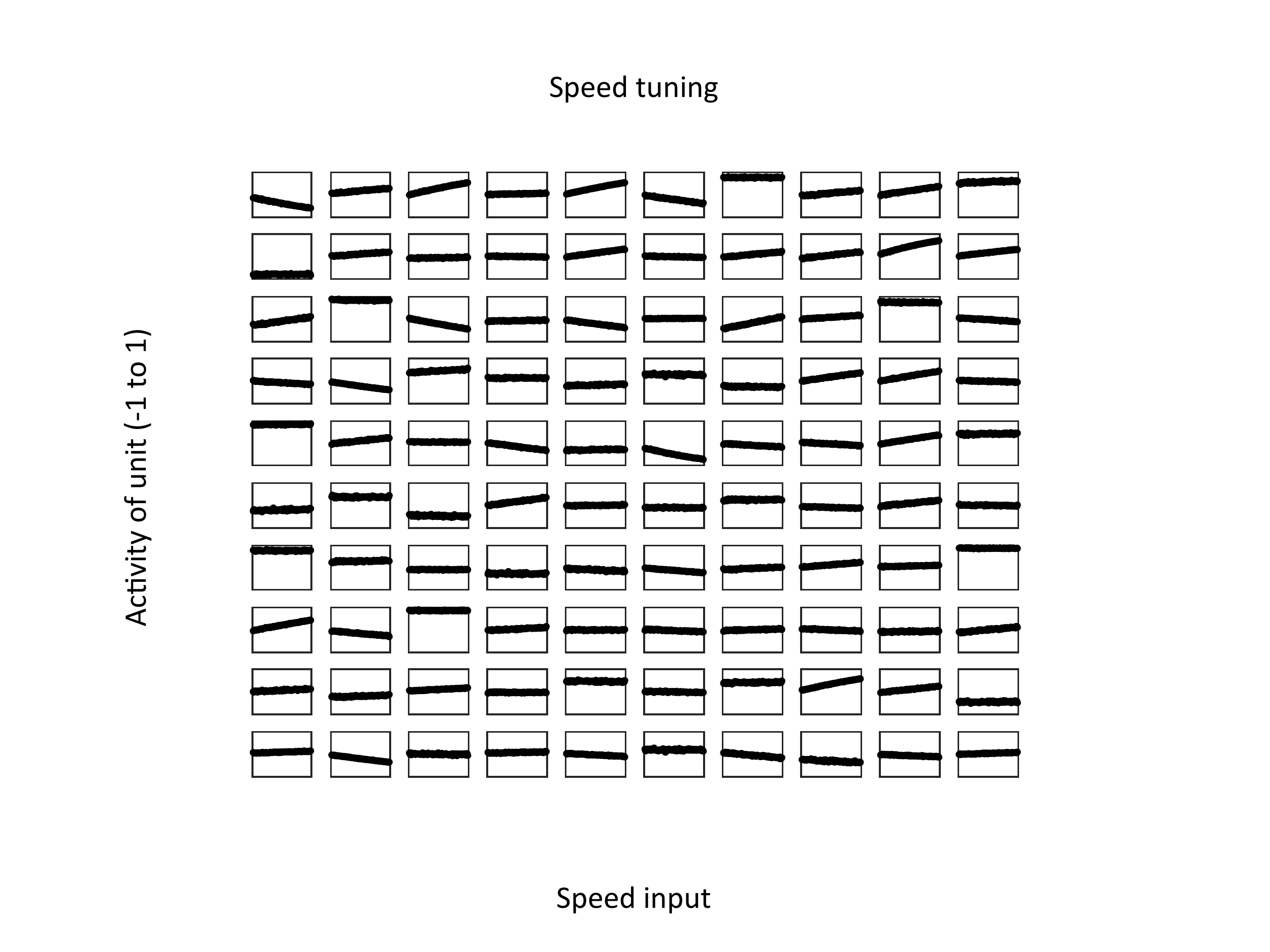}
\end{figure}

\newpage
\section{Hexagonal environment}
\begin{figure}[h]
 \centering
\includegraphics[keepaspectratio,width=0.48\linewidth]{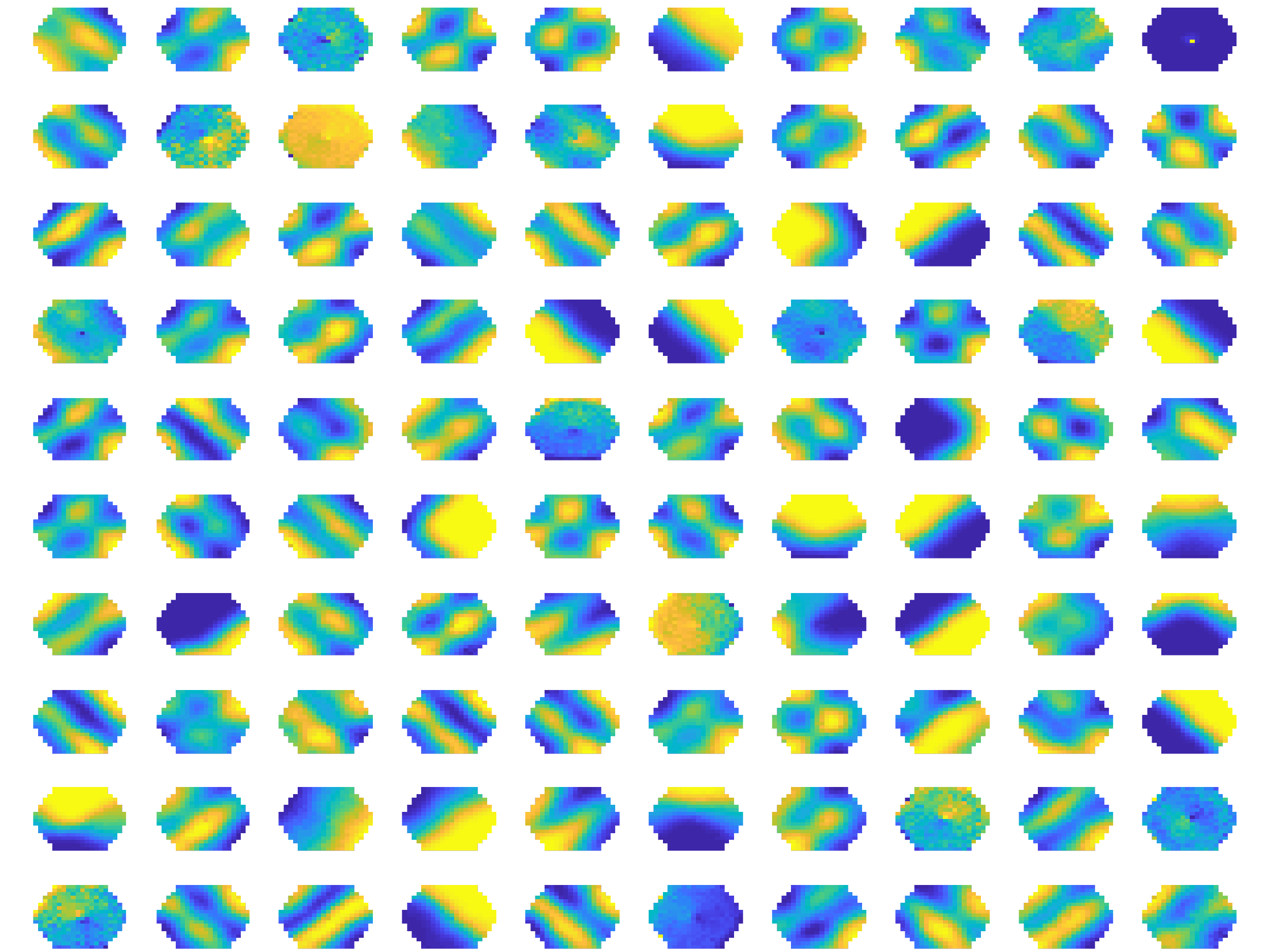}
\end{figure}
\begin{figure}[h]
 \centering
\includegraphics[keepaspectratio,width=0.68\linewidth]{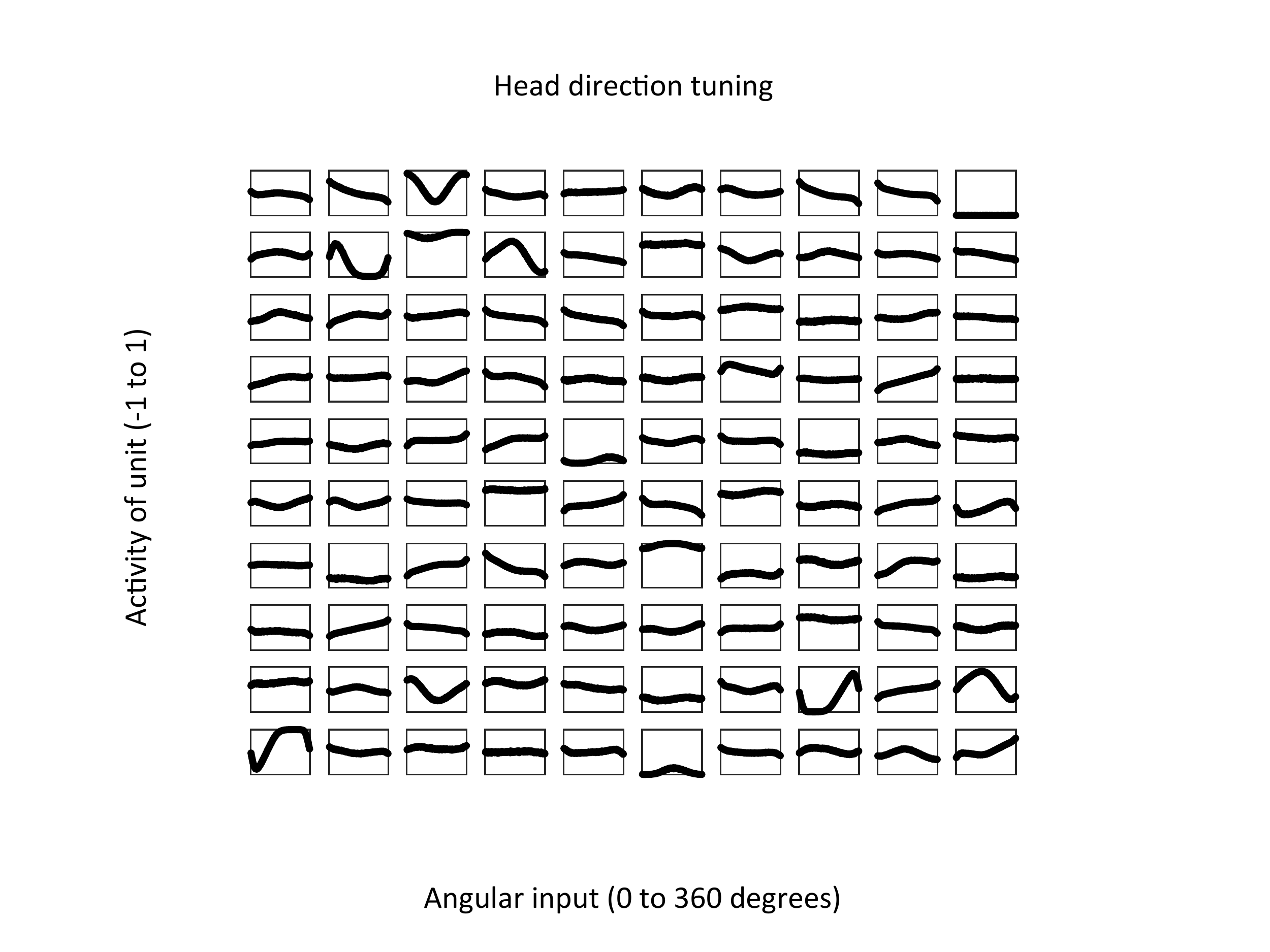}
\end{figure}
\begin{figure}[h]
 \centering
\includegraphics[keepaspectratio,width=0.68\linewidth]{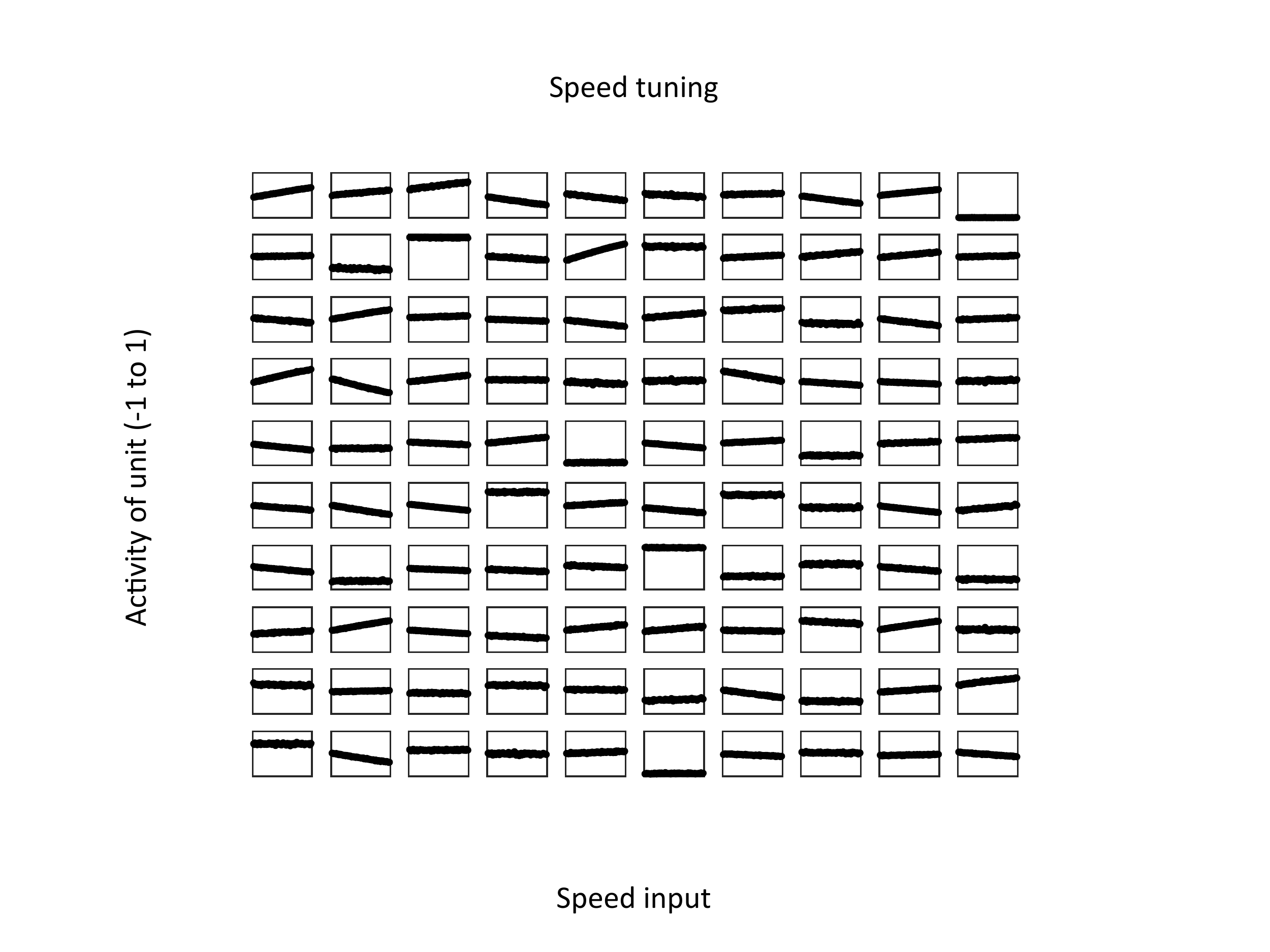}
\end{figure}

\newpage
\section{Relation between speed, direction, and spatial selectivity}
To quantify the speed selectivity of each unit we first fit a line to the tuning curve of unit activity as a function of speed. The speed selectivity is the absolute value of the slope. If the unit activity is not modulated by speed then the speed selectivity is 0. 
To quantify the direction selectivity of each unit we calculated the average unit activity as a function of direction input and then took the maximum minus minimum of this tuning curve. If the unit activity is not modulated by direction then the direction selectivity is 0.
To quantify the spatial selectivity we used lifetime sparseness~\citep{Willmore2001}. If the unit activity is not modulated by spatial location then the spatial selectivity is 0. Each dot in the figures below show the selectivity for a single unit.
\begin{figure}[h]
 \centering
\includegraphics[trim={0cm 9.5cm 0cm 0cm},clip,keepaspectratio,width=1.0\linewidth]{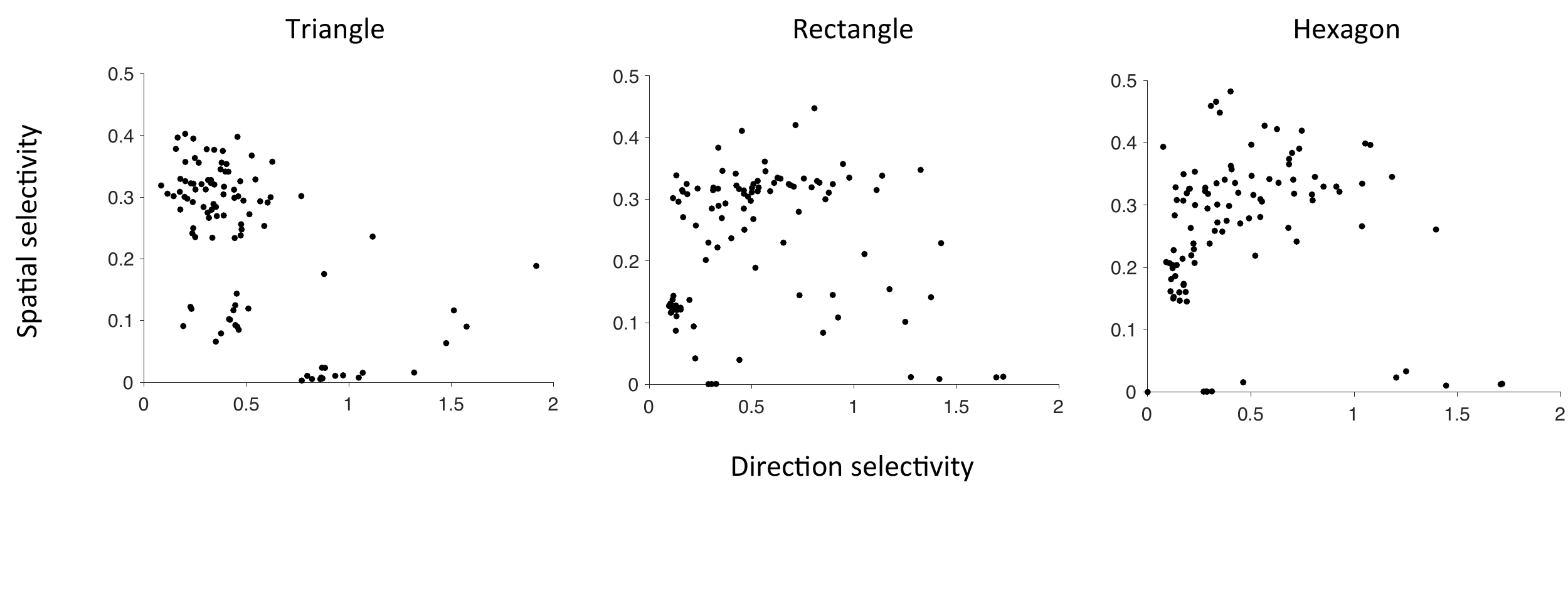}
\end{figure}

\begin{figure}[h]
 \centering
\includegraphics[trim={0cm 9.5cm 0cm 0cm},clip,keepaspectratio,width=1.0\linewidth]{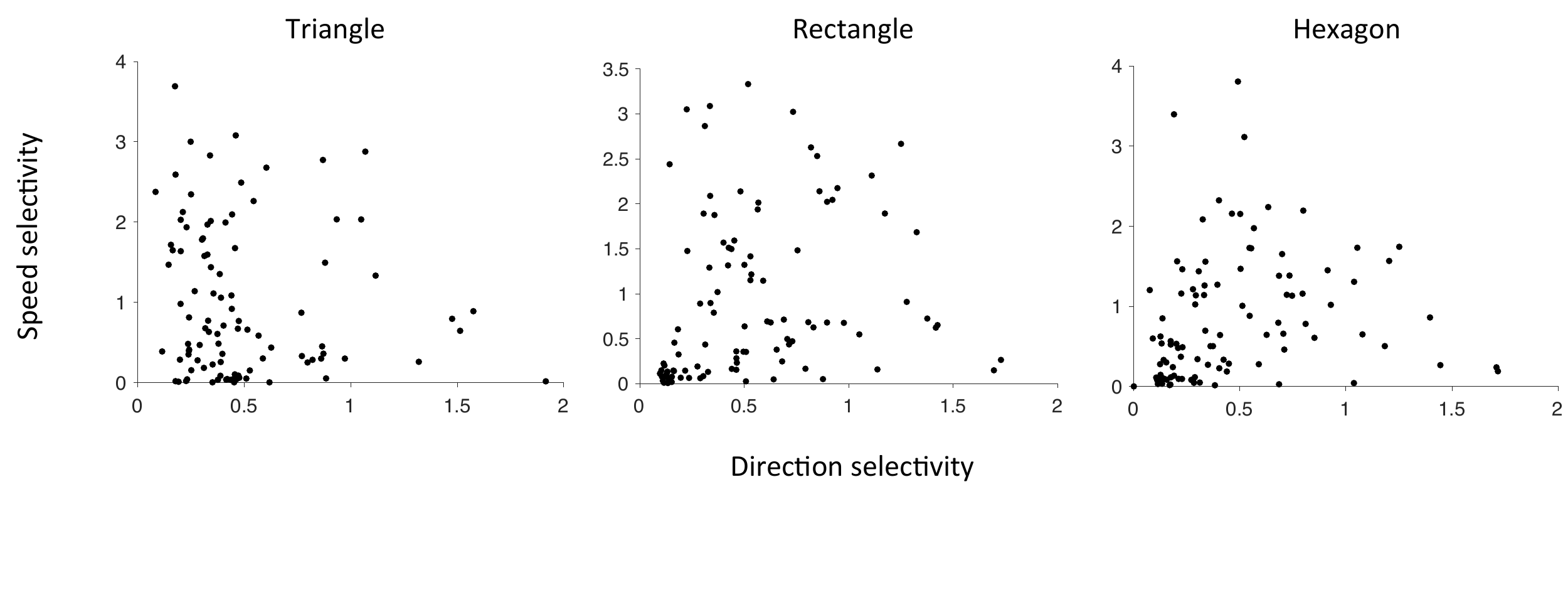}
\end{figure}

\begin{figure}[h]
 \centering
\includegraphics[trim={0cm 9.5cm 0cm 0cm},clip,keepaspectratio,width=1.0\linewidth]{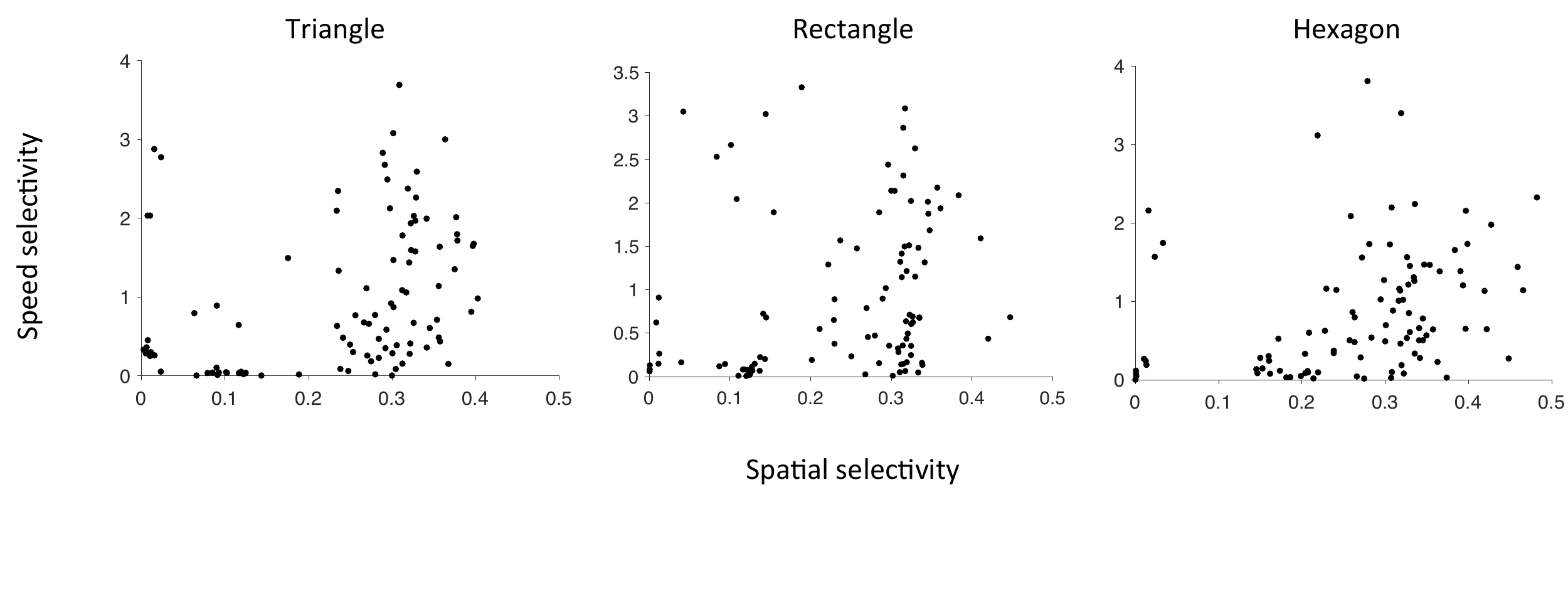}
\end{figure}

\newpage
\section{Additional training details}
During training we tried to balance all three terms we were minimizing ($E$, $R_{L2}$, and $R_{FR}$) so no single term was neglected or dominated.  At the beginning of training we weighted the regularization term $R_{L2}$ to be equal to the error function $E$ and then decreased the weighting on $R_{L2}$ according to the schedule used by~\cite{Martens2011}. We adaptively adjusted the weighting on $R_{FR}$, starting from an initial value of $E/10$ and enforcing an upper bound of $E/3$ as training progressed. We found this training procedure improved training performance and led to more interesting representations.

\end{document}